\documentclass[aps,twocolumn,floatfix,prl]{revtex4}
\usepackage{graphicx,subfigure,amsmath,bbm}

\def\braket#1{\mathinner{\langle{#1}\rangle}}
\def\bra#1{\left\langle#1\right|}
\def\ket#1{\left|#1\right\rangle}

\begin{document}

\title{Pauli Spin Blockade in the Presence of Strong Spin-Orbit Coupling}
\date{\today}

\author{J.\ Danon}
\author{Yu.~V.\ Nazarov}
\affiliation{Kavli Institute of NanoScience, Delft University of Technology, 2628 CJ Delft, The Netherlands}

%\pacs{??}
%{03.67.Lx,73.63.Kv,76.30.-v}
%73.23.Hk Coulomb blockade; single-electron tunneling
%03.67.Lx Quantum computation
%03.67.-a Quantum information
%76.30.-v Electron paramagnetic resonance and relaxation
%73.63.Kv Quantum dots

\begin{abstract}
We study electron transport in a double quantum dot in the Pauli spin blockade regime, in the presence of strong spin-orbit coupling. The effect of spin-orbit coupling is incorporated into a modified interdot tunnel coupling. We elucidate the role of the external magnetic field, the nuclear fields in the dots, and spin relaxation. We find qualitative agreement with experimental observations, and we propose a way to extend the range of magnetic fields in which blockade can be observed.
\end{abstract}

\maketitle

%\section{Introduction}

Blockade phenomena, whereby strong interactions between single particles affect the global transport or excitation properties of a system, are widely used to control and detect quantum states of single particles. In single electron transistors, the electrostatic interaction between electrons can block the current flow~\cite{fulton:prl}, thereby enabling precise control over the number of charges on the transistor~\cite{ashoori:prl}.
%Analogous blockade phenomena were observed using cold atoms in optical dipole traps~\cite{schlosser:nature} and photons in non-linear optical cavities~\cite{birnbaum:nature}.
%
In semiconductor quantum dots, the Pauli exclusion principle can lead to a spin-selective blockade~\cite{ono:science}, which has proven to be a powerful tool for read-out of the spin degree of freedom of single electrons~\cite{frank:nature,katja:science,petta:science,frank:science,johnson:nature}.

\begin{figure}[b]
\includegraphics[width=8.5cm]{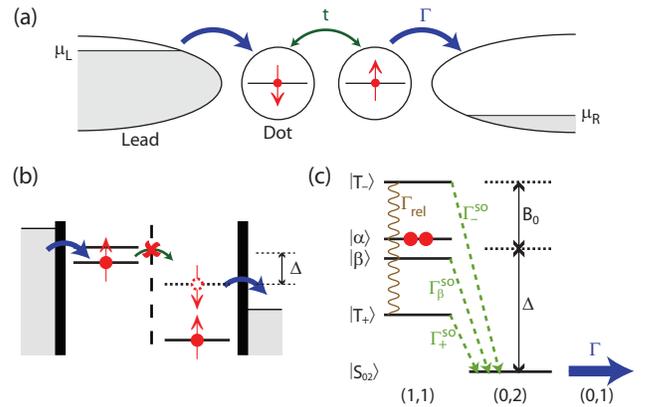}
\caption{Double quantum dot in the Pauli spin blockade regime. (a) The double dot is coupled to two leads. Due to a voltage bias, electrons can only run from the left to the right lead. (b) Energy diagram assuming {\em spin-conserving} interdot coupling. The only accessible $(0,2)$ state is a spin singlet: all $(1,1)$ triplet states are not coupled to the $(0,2)$ state and the current is blocked. (c) Energy levels and transition rates assuming {\em non-spin-conserving} interdot coupling. We consider the `high'-field limit and neglect the effects of the nuclear fields. Then three of the four $(1,1)$ states can decay, leaving only one spin blockaded state $\ket{\alpha}$. Isotropic spin relaxation $\sim \Gamma_\text{rel}$ causes transitions between all $(1,1)$ states.} \label{fig:fig1}
\end{figure}

In this spin blockade regime, a double quantum dot is tuned such that current involves the transport cycle $(0,1) \to (1,1) \to (0,2) \to (0,1)$, $(n,m)$ denoting a charge state with $n(m)$ excess electrons in the left(right) dot (see Fig.\ \ref{fig:fig1}(a)). Since the only accessible $(0,2)$ state is a spin singlet, the current is blocked as soon as the system enters a $(1,1)$ triplet state (Fig.\ \ref{fig:fig1}(b)): transport is then due to spin relaxation processes, possibly including interaction with the nuclear fields~\cite{jouravlev:prl}. This blockade has been used in GaAs quantum dots to detect coherent rotations of single electron spins~\cite{frank:nature,katja:science}, coherent rotations of two-electron spin states~\cite{petta:science}, and mixing of two-electron spin states due to hyperfine interaction with nuclear spins~\cite{johnson:nature,frank:science}.

Motivated by a possibly large increase of efficiency of magnetic and electric control over the spin states~\cite{flindt:prl,golovach:prb}, also quantum dots in host materials with a relatively large $g$-factor and strong spin-orbit interaction are being investigated.
%It is believed that a relatively large $g$-factor and strong spin-orbit interaction significantly enhances the efficiency of the magnetic and electric control of spin states in quantum dots~\cite{flindt:prl,golovach:prb}. This motivated intensive research on quantum dots in these materials.
Very recently, Pauli spin blockade has been demonstrated in a double quantum dot defined by top gates along an InAs nanowire~\cite{pfund:prl,pfund:pe}.
However, as compared to GaAs, spin blockade in InAs nanowire quantum dots seems to be destroyed by the strong spin-orbit coupling: significant spin blockade has been only observed at very small external magnetic fields ($\lesssim$ 10 mT~\cite{pfund:prl}).
%If this would be a general property of these quantum dots, it would restrict the range of applications significantly.
An important question is whether there exists a way to extend this interval of magnetic fields. To answer that question, one first has to understand the physical mechanism behind the lifting of the blockade.

In this work we study Pauli spin blockade in the presence of strong spin-orbit mixing. We show that the only way spin-orbit coupling interferes with electron transport through a double dot is by introducing non-spin-conserving tunneling elements between the dots. This yields coupling of the $(1,1)$ triplet states to the outgoing $(0,2)$ singlet, thereby lifting the spin blockade.
%seeming to lift the blockade completely.
However, for sufficiently small external magnetic fields this does not happen. If the $(1,1)$ states are not split apart by a large Zeeman energy, they will rearrange to one coupled, decaying state and three blocked states. When the external field $B_0$ is increased, it couples the blocked states to the decaying state. As soon as this field induced decay grows larger than the other escape rates (i.e.\ $B_0^2 \Gamma/t^2 > \Gamma_\text{rel}$, where $\Gamma$ is the decay rate of the $(0,2)$ singlet, $t$ the strength of the tunnel coupling, and $\Gamma_\text{rel}$ the spin relaxation rate~\cite{endnote1}) the blockade is lifted. Therefore, the current exhibits a dip at small fields.

The presence of two random nuclear fields in the dots (of typical magnitude $K \sim 1$~mT) complicates matters since it adds another dimension to the parameter space. We distinguish two cases: If the nuclear fields are small compared to $t^2/\Gamma$, they just provide an alternative way to escape spin blockade, which may compete with spin relaxation. There is still a dip at small magnetic fields, and the current and width of the dip are determined by the maximum of $ \Gamma_\text{rel}$ and $K^2 \Gamma/t^2$. In the second case, $K \gg t^2/\Gamma$, the current may exhibit either a peak or a dip, depending on 
the strength and orientation of the spin-orbit mixing. If there is a peak in this regime, the cross-over from dip to peak takes place at $K\sim t^2/\Gamma$.
% they provide a decay process competing with $\Gamma_\text{rel}$. 
% Current then scales as $I/e \sim \text{max}\{ \Gamma_\text{rel}, K^2 \Gamma/t^2 \}$, and the transition from low to high current is observed at $B_0^2 \Gamma/t^2 \sim \text{max}\{ \Gamma_\text{rel}, K^2 \Gamma/t^2 \}$. If, on the other hand the nuclear fields are larger than $t^2/\Gamma$, current will scale as $I/e \sim t^2/\Gamma$. Around zero external field now either a peak or dip in the current can be observed, this feature disappears at $B \sim K$.

%These qualitative results agree with experimetal observations~\cite{pfund:prl,churchill:arxiv}, and elucidate the role of spin-orbit coupling, spin relaxation, and nuclear fields in the spin blockade regime. Having understood the undelying physical mechanism, we propose a practical way to extend the magnetic field interval of spin blockade. With a freely rotatable magnet, one can always align $B_0$ in such a way that two of the three triplet states will be blocked. 

%\section{Model and Approach}

%In the Pauli spin blockade regime, a double quantum dot is tuned such that the only possible transport cycle is $(1,1) \to (0,2) \to (0,1) \to (1,1)$, where $(n,m)$ denotes the charge state with $n(m)$ excess electrons in the left(right) dot. The only $(0,2)$ state which is energetically accessible is a spin singlet, implying that, if the interdot coupling is spin-conserving, further flow of electrons is blocked as soon as the system enters a $(1,1)$ triplet state (see Fig.\ \ref{fig:fig1}).

Let us now turn to our model. We describe the relative detuning of the $(1,1)$ states and the $(0,2)$ states by the Hamiltonian $\hat H_e = -\Delta \ket{S_{02}}\bra{S_{02}}$, where $\ket{S_{02}}$ denotes the $(0,2)$ spin singlet state.
The energies of the four $(1,1)$ states are further split by the magnetic fields acting on the electron spins,
$\hat H_m =  B_0 (\hat S^z_L+\hat S^z_R) +  \vec{K}_L \cdot \hat{\vec{S}}_L + \vec{K}_R \cdot \hat{\vec{S}}_R$,
where  $\hat{\vec{S}}_{L(R)}$ is the electron spin operator in the left(right) dot (for InAs nanostructures $g \sim$ 7~\cite{pfund:prb}). We chose the $z$-axis along $\vec B_0$, and  included two randomly oriented effective nuclear fields $\vec{K}_{L,R}$ resulting from the hyperfine coupling of the electron spin in each dot to 
%$N \sim 10^5$ nuclear spins~\cite{pfund:prb}. The typical magnitude of the random nuclear fields is $\propto 1/\sqrt{N}\sim 0.6$ $\mu$eV.
$N$ nuclear spins (in InAs quantum dots $N \sim 10^5$~\cite{pfund:prb}, yielding a typical magnitude $K \propto 1/\sqrt{N}\sim 0.6$ $\mu$eV).
% 0.6 ueV = 1.5 mT with g = 7
We treat the nuclear fields classically, disregarding
feedback of the electron spin dynamics which
could lead to dynamical nuclear spin polarization 
\cite{danon:arxiv}.

Let us now analyze the possible effects of spin-orbit coupling. 
(i) It can mix up the spin and orbital structure of the electron states.
The resulting states will remain Kramers doublets, thus giving no
qualitative difference with respect to the common spin up and down doublets.
(ii) The mixing also renormalizes the $g$-factor that defines 
the splitting of the doublets in a magnetic field.
This, however, is not seen provided we measure $B_0$ in units of energy.
(iii) The coupling also can facilitate spin relaxation~\cite{khaetskii:prb},
%$\Gamma_\text{rel}$
but this is no qualitative change either.
Some of these aspects have been investigated in~\cite{sodqd}.

The only place where strong spin-orbit interaction leads to a qualitative
change is the tunnel coupling between the dots. 
It provides a finite 
overlap of states differing in index of the Kramers doublet
(in further discussion we refer to this index as to `spin'),
this allowing for a very compact model 
incorporating the interaction. The most general non-`spin'-conserving 
tunneling Hamiltonian for two doublet electrons in a left and right state reads
$\hat H_t = \sum_{\alpha,\beta} \left\{ t^L_{\alpha\beta} \hat a^\dagger_{L\alpha}\hat a_{R\beta} + t^R_{\alpha\beta} \hat a^\dagger_{R\alpha}\hat a_{L\beta} \right\}$,
$\alpha , \beta$ being the spin indices, 
$\hat a^\dagger_{L(R)}$ and $\hat a_{L(R)}$ the electron 
creation and annihilation operators in the left(right) state, 
and $t^{L,R}$ coupling matrices.
We impose conditions of hermiticity and time-reversibility on $\hat H_t$ 
and concentrate on the matrix elements between the $(1,1)$ states and $\ket{S_{02}}$
in our double dot setup. In the convenient basis of  
orthonormal unpolarized triplet states $\ket{T_{x,y}} 
\equiv i^{1/2\mp 1/2}\{\ket{T_-}\mp \ket{T_+}\} /\sqrt{2}$, 
%$\ket{T_y} \equiv i\{ \ket{T_-}+\ket{T_+}\} /\sqrt{2}$, 
$\ket{T_z} \equiv \ket{T_0}$, and the $(1,1)$ singlet $\ket{S}$, this 
Hamiltonian reads
\begin{equation}
\hat H_t = i \vec t \cdot \vec{\ket{T}}\bra{S_{02}} + t_0 \ket{S}\bra{S_{02}} + \text{h.c.},
\label{eq:hamt2}
\end{equation}
with $\vec{\ket{T}} \equiv \{ \ket{T_x}, \ket{T_y}, \ket{T_z} \}$.
The model therefore adds a 3-vector of new coupling parameters, $\vec t = \{t_x, t_y, t_z \}$, 
to the usual spin-conserving $t_0$,
the vector being a `real' vector with respect to coordinate transformations.
If the energy scale of spin-orbit interaction is larger or
comparable to the energy distance between the levels in the dot 
(which is believed to be the case in InAs structures),
the mixing of the doublet components is of the order of $1$.
Therefore we assume that all four coupling parameters are 
generally of the same order of magnitude $t_{0,x,y,z} \sim t$.
As the structure of the localized
electron wave functions is very much dependent on the nanostructure
design and its inevitable imperfections, the direction of $\vec{t}$
is hard to predict: therefore we consider arbitrary directions.

We describe the electron dynamics with an evolution equation 
for the density matrix~\cite{jouravlev:prl}.
Next to the Hamiltonian terms, we complement the equation
with (i) the rates $\sim \Gamma$ 
describing the decay of $\ket{S_{02}}$ and the refill to a $(1,1)$ state,
and (ii) a small electron spin relaxation rate $\Gamma_\text{rel} \ll \Gamma$.
The full evolution of the electron density matrix then can be written as
\begin{equation}
\frac{d\hat \rho}{dt} = -i
[ \hat H_e+\hat H_m+\hat H_t,\hat \rho ] + \mathbf{\Gamma}\hat \rho + \mathbf{\Gamma}_\text{\!rel} \, \hat \rho.
\label{eq:evol}
\end{equation}
Experimentally, the temperature exceeds the Zeeman energy~\cite{pfund:prl},
allowing us to assume isotropic spin relaxation: each $(1,1)$ state
will transit to any of the other $(1,1)$ states with a rate $\Gamma_\text{rel}/3$.
Explicitly, we use $\mathbf{\Gamma}_\text{\!rel} \, \hat \rho = -\Gamma_\text{rel} \hat \rho + \frac{1}{6} \Gamma_\text{rel} \sum_{\alpha,d} \hat \sigma_d^\alpha \hat \rho\hat \sigma_d^\alpha$, $\hat \sigma^\alpha_{L(R)}$ being the Pauli matrices in the left(right) dot.

Motivated by experimental work, we assume that 
the decay rate $\Gamma$ of $\ket{S_{02}}$ 
is by far the largest frequency scale in (\ref{eq:evol}),
i.e.\ $\Gamma \gg B_0,K,t,\Gamma_\text{rel}$ (in principle $\Gamma$ can be
comparable with the detuning $\Delta$).
Under this assumption, we separate the time scales and 
derive the effective evolution equation for the density matrix in 
the $(1,1)$ subspace
\begin{equation}
\frac{d\hat \rho}{dt} = -i
[\hat H_m + \hat H_t',\hat \rho ] - \mathbf{G}^\text{out}\hat \rho
+\mathbf{G}^\text{in}\hat \rho
+ \mathbf{\Gamma}_\text{\!rel} \, \hat \rho.
\label{eq:evol2}
\end{equation}
The decay and refill terms are now incorporated into
%the superoperators
\begin{equation}
\begin{split}
G^\text{out}_{kl,mn} & = 2
\{ \delta_{km}T_{n2}T_{2l} +\delta_{ln}T_{k2}T_{2m} \}\Gamma/(\Gamma^2+4\Delta^2) \\
G^\text{in}_{kl,mn} &= \delta_{kl}T_{n2}T_{2m}\Gamma/(\Gamma^2+4\Delta^2),
\end{split}
\end{equation}
where $T_{a2} \equiv \braket{a|\hat H_t |S_{02}}$.
The coupling between the dots 
gives also rise to an exchange Hamiltonian
%$\hat H_t' = \frac{4\Delta}{\Gamma^2+4\Delta^2} \hat H_t\ket{S_{02}}\bra{S_{02}}\hat H_t$, 
$(H_t')_{ij} = 4\Delta /(\Gamma^2+4\Delta^2) T_{i2} T_{2j}$, with
$H_t' \sim G^\text{out}$ provided $\Gamma \sim \Delta$.
The diagonal elements of $\mathbf{G}^\text{out}$ 
give us the decay rates:
If we consider $\ket{T_\pm}$ and $\ket{T_0}$, 
the three triplet states split by an external magnetic field, 
we find $\Gamma^\text{so}_\pm \equiv G^\text{out}_{\pm \pm,\pm \pm} = 
2\Gamma (t_x^2+t_y^2)/(\Gamma^2+4\Delta^2)$ and
$\Gamma^\text{so}_0 \equiv G^\text{out}_{0 0,0 0} = 
4\Gamma t_z^2/(\Gamma^2+4\Delta^2)$,
all of which are $\sim \Gamma^\text{so} \sim t^2/\Gamma$.

Let us neglect for a moment the nuclear fields,
and focus on zero detuning, $\Delta = 0$.
%and assume $\Gamma_{rel}\ll t^2/\Gamma$. 
This allows us to grasp qualitatively the peculiarities 
of the spin blockade lifting, determined by
competition between the Hamiltonian ($\sim B_0$) and dissipative
terms ($\sim t^2/\Gamma, \Gamma_\text{rel}$) in Eq.\ \ref{eq:evol2}. 

At sufficiently large fields, 
%$B_0 \gg t^2/\Gamma$, 
the basis states $\ket{T_0}$ 
and $\ket{S}$ are aligned in energy.
The spin-orbit modulated tunnel coupling then sets the difference 
between these states, which is best seen in a basis that mixes the 
states, $\ket{\alpha} \equiv \{ t_0 \ket{T_0} + it_z\ket{S}\}/\sqrt{t_0^2+t_z^2}$ and $\ket{\beta} \equiv \{ it_z\ket{T_0} + t_0\ket{S} \}/\sqrt{t_0^2+t_z^2}$. Now $\ket{\alpha}$ is a blocked state, i.e.\ 
$G^\text{out}_{\alpha\alpha,\alpha\alpha}=0$,
while $\ket{\beta}$ decays with 
an effective rate $\Gamma^\text{so}_\beta \equiv G^\text{out}_{\beta\beta,\beta\beta} = 
4\Gamma(t_0^2+t_z^2)/(\Gamma^2+4\Delta^2)$. 
In Fig.\ \ref{fig:fig1}(c) we give the energy levels of the five states and 
all transition rates in the limit of `large' external fields. 
It is clear that the system will spend most time 
in the state $\ket{\alpha}$. 
The current is determined by the spin-relaxation decay rate of this state
to any unblocked state, 
$3\Gamma_{\text{rel}}/3 = \Gamma_{\text{rel}}$. Let us 
note that if $n_b$ states out of $n$ states are blocked,
such a decay produces on average $n/n_b$ electrons tunneling 
to the outgoing lead before the system is recaptured in a blocked state.
Therefore, the current is $I/e = 4\Gamma_\text{rel}$.

This picture holds until the decay rates of the three non-blocked states
become comparable with $\Gamma_\text{rel}$,
which takes place at $B_0 \sim \sqrt{\Gamma^\text{so}\Gamma_\text{rel}}$.
To understand this, let us start with considering
the opposite limit, $B_0 \ll \sqrt{\Gamma^\text{so} \Gamma_\text{rel}}$.
In this case all four $(1,1)$ states are almost aligned in energy, 
and the instructive basis to work in is the one spanned by 
a single decaying state $\ket{m} \equiv \{i\vec t \cdot \vec{\ket{T}} + t_0\ket{S} \}/\sqrt{|\vec t|^2 + t_0^2}$, 
and three orthonormal states $\ket{1}$, $\ket{2}$ and $\ket{3}$
that are not coupled to $\ket{S_{02}}$.
At $B_0 = 0$ three of the four states are blocked, and
spin relaxation to the unblocked state
proceeds with a rate $\Gamma_\text{rel}/3$. A relaxation process produces
on average $n/n_b = 4/3$ electron transfers, so that the total current 
is reduced by a factor of $9$ in comparison with the `high'-field case,
$I/e = \frac{4}{9}\Gamma_\text{rel}$. This factor of 9 agrees 
remarkably well with experimental observations 
(see Fig.\ 2b in Ref.\ \cite{pfund:prl}).

\begin{figure}[b]
\includegraphics[width=8.5cm]{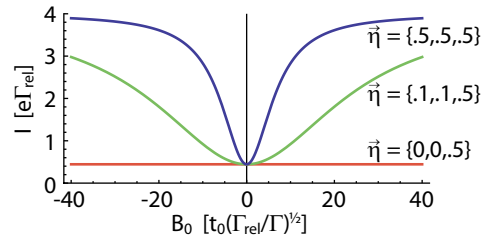}
\caption{Current as a function of $B_0$, at $\Delta = 0$, and neglecting the nuclear fields. Around zero field a dip is observed, its width depends on the magnitude and orientation of $\vec \eta$.}\label{fig:fig3}
\end{figure}

We now add a finite external field $B_0$ to this picture. Since 
$\vec t$ is generally not parallel to $B_0$, the external field 
will split the states $\ket{1}$, $\ket{2}$ and $\ket{3}$ 
in energy and mix two of them with the decaying state $\ket{m}$. 
This mixing results in an effective decay rate $\sim B_0^2 
/\Gamma^\text{so}$, which may compete 
with the spin relaxation rate $\Gamma_\text{rel}$.
At $B_0 \sim \sqrt{\Gamma^\text{so}\Gamma_\text{rel}}$,
we cross over to the `high'-field regime described above,
where only one blocked state is left. Therefore, the current
exhibits a dip (suppression by a factor 9) 
around zero field with a width estimated as 
$\sqrt{\Gamma^\text{so}\Gamma_\text{rel}}$ (Fig.\ \ref{fig:fig3}).

Let us now include the effects of the nuclear fields $\vec{K}_{L,R}$
on a qualitative level. If the fields are small compared to the scale 
$t^2 /\Gamma$, their only relevant effect is to mix 
the states described above. This mixing creates a new possibility for 
decay of the blocked states, characterized by a
rate $\Gamma_N \sim K^2/\Gamma^\text{so}$. This rate
may compete with spin relaxation $\sim \Gamma_\text{rel}$, and
could cause the current to scale with $\Gamma_N$ 
and the width of the dip with $K$.
In the opposite limit, $K \gg t^2 /\Gamma$, 
the nuclear fields dominate the energy scales 
and separation of the $(1,1)$ states at $B_0 \lesssim K$. 
Then, generally all four states are coupled to $\ket{S_{02}}$ 
on equal footing and the spin blockade is lifted.
Qualitatively, this situation is similar to that without 
spin-orbit interaction (see Eqs 10-12
in~\cite{jouravlev:prl}). {\em Without} spin-orbit interaction,
an increase of magnetic field leads to blocking of
two triplet states, resulting in a current peak at zero field.
{\em With} spin-orbit interaction, $t_{x,y}$ still couple the 
split-off triplets to the decaying state. Depending on
the strength and orientation of $\vec t$,
the current in the limit of `high' fields can be either smaller or larger
than that at $B_0=0$, so we expect either peak or dip.
If it is a peak, the transition from
peak to dip is expected at $K \sim \Gamma^\text{so}$, that is, at $t \sim
\sqrt{K\Gamma}$. 
Indeed, such a transition has been observed 
upon varying the magnitude of the
tunnel coupling (Fig.\ 2 in Ref.\ \cite{pfund:prl}).
If we assume that $K \sim 1.5$~mT and
associate the level broadening observed ($\sim 100$~$\mu$eV) with
$\Gamma$, we estimate $t \sim 8$~$\mu$eV, which agrees with the range of
coupling energies mentioned in~\cite{pfund:prl}.

Let us now support the qualitative arguments given above with explicit analytical and numerical solutions. The current through the double dot is evaluated as $I/e = \rho_{22} \Gamma$, $\rho_{22}$ being the steady-state probability 
to be in $\ket{S_{02}}$, as obtained from solving Eq.\ \ref{eq:evol}.
We give an analytical solution for $\Delta = 0$, neglecting the nuclear fields,
and expressing the answer in terms of the dimensionless parameter
$\vec t/t_0 = \vec \eta$. Under these assumptions, we find
\begin{equation}
I = I_\text{max}\left( 1-\frac{8}{9} \frac{B^2_c}{B^2+B^2_c}\right),
\label{eq:curr}
\end{equation}
with $B_c =  2\sqrt{2} (1+|\vec \eta|^2)(\eta_x^2+\eta_y^2)^{-1/2}t_0\sqrt{\Gamma_\text{rel}/\Gamma}$
and $I_\text{max} = 4e\Gamma_\text{rel}$.
The current exhibits a Lorentzian-shaped dip (see Fig.\ \ref{fig:fig3}, 
compare with Fig.\ 2b in Ref.\ \cite{pfund:prl}).
The width $B_c$ and the limits at low and `high' fields agree with the
qualitative estimations given above.

%In the limit where $K \gg t^2/\Gamma_\text{out}, \Gamma_\text{rel}$
%$I/e\Gamma_\text{out} = \frac{1}{8} \{ \xi_+ - (\xi_- \vec n_L\cdot \vec n_R + 2 (\vec t\cdot \vec n_L)(\vec t\cdot \vec n_R) + 2 t_0 \vec t\cdot(\vec n_L \times \vec n_R))^2/\xi_+\}$

\begin{figure}[t]
\includegraphics[width=8.5cm]{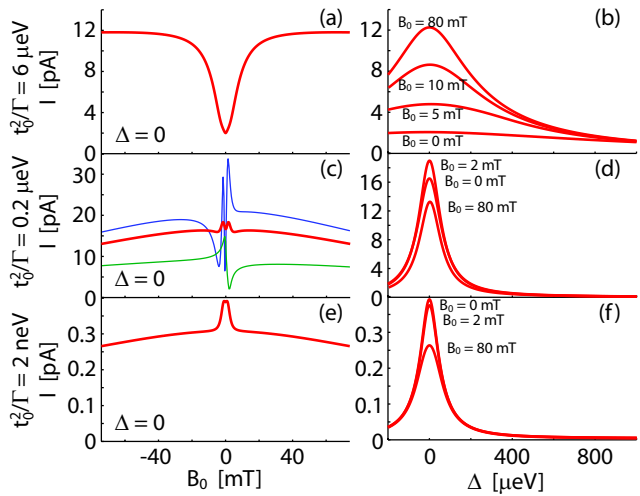}
\caption{The current $I = e\rho_{22} \Gamma$ for (a,b) large, (c,d) intermediate, and (e,f) small tunnel coupling. The dip observed around zero field (a) disappears when $t_0^2/\Gamma \sim K$ (c) and evolves into a peak for even smaller tunnel coupling (e).}\label{fig:fig4}
\end{figure}
To include the effect of the two nuclear fields, 
we compute steady-state solutions of (\ref{eq:evol})
% at a given configuration of nuclear fields $\vec{K}_{L,R}$ 
and average over many configurations of $\vec{K}_{L,R}$~\cite{jouravlev:prl}.
In Fig.\ \ref{fig:fig4} we present the resulting current 
versus magnetic field and detuning
for three different regimes. To produce the plots 
we turned to concrete values of the parameters, setting 
$\Gamma = $ 0.1 meV, $\Gamma_\text{rel} = $ 1~MHz, 
$\vec \eta = 0.25\times\{1,1,1\}$. We averaged over 5000 configurations of $\vec{K}_{L,R}$, randomly 
sampled from a normal distribution with a r.m.s.\ of 0.4 $\mu$eV.

In Fig.\ \ref{fig:fig4}(a) and (b) we assumed large tunnel coupling, 
$t_0^2/\Gamma = 6$~$\mu$eV so that $ K\Gamma/t_0^2 = 0.07$ 
is small. In (a) we plot the current at $\Delta = 0$, while in (b) 
we plot it versus detuning for different fixed $B_0$. 
We observe in (a) a Lorentzian-like dip in the current at $B_0 = 0$.
While it looks similar to the plots in Fig.\ \ref{fig:fig3},
the width is determined by the nuclear fields since $K \gg \Gamma_\text{rel}$.
The curve can be accurately fit with the Lorentzian (\ref{eq:curr}),
giving $B_c = 7.4~K$ and
% = 7.44 and K = 1 mT
$I_\text{max} = 0.62~K^2\Gamma/t_0$.
% 11.97 pA = 0.0493 ueV
% K^2\Gamma/t_0^2 = 0.08 ueV
%and a broadening of the current peak around $\Delta = 0$. 
Fig.\ \ref{fig:fig4}(b) illustrates the unusual broadening of the resonant peak 
with respect to its natural width determined by $\Gamma$.
The width in this case scales as $\sim t_0^2/K$ and
is determined by competition of $\Gamma^\text{so}$ and $\Gamma_N$. 
These plots qualitatively agree with data presented 
in Fig.\ 2b in Ref.\ \cite{pfund:prl}.
%we see that the dip and the broadening both agree qualitatively with the experimental data. 
%The asymmetry in the current peak around zero detuning observed in experiment is due to inelastic tunneling processes which we did not include in our model
%
In Fig.\ \ref{fig:fig4}(c) and (d) we present the same plots,
for smaller tunnel coupling, $t_0^2/\Gamma = 0.2$~$\mu$eV $ = 0.5~K$. 
We included in plot (c) 
the curves for two random nuclear field configurations: 
It is clear that the current strongly depends 
on $\vec{K}_{L,R}$, which agrees with our expectation 
that in the regime $\Gamma_\text{rel} < \Gamma_N$ 
the current $I \propto \Gamma_N \propto K^2$. 
Remarkably, averaging over many configurations smooths 
the sharp features at small $B_0$ (c.f.\ \cite{jouravlev:prl}). 
Plots (d) exhibit no broadening with respect to $\Gamma$, 
in correspondence with Fig.\ 2a of Ref.\ \cite{pfund:prl}.
In Fig.\ \ref{fig:fig4}(e) and (f) we again made the same plots 
for yet smaller tunnel coupling, $t_0^2/\Gamma = 2$~neV 
$ \ll K$. Since the nuclear fields now dominate the splitting
of the $(1,1)$ states, we see a peak comparable to the one in 
Fig.\ 4 of Ref.\ \cite{jouravlev:prl} surmounting a finite background 
current due to spin-orbit decay of the split-off triplets. 

We expect our results to hold for any quantum dot system with strong spin-orbit interaction. Indeed, recent experiments on quantum dots in carbon nanotubes in the spin blockade regime~\cite{churchill:arxiv} 
display the very same specific features, as e.g.\ a zero-field dip in the current.

Now that we understand the origin of the lifting of spin blockade, we also propose a way to extend the blockade region. If one would have a freely rotatable magnet as source of the field $B_0$, one would observe a large increase in width of the blockade region as soon as $\vec B_0$ and $\vec t$ are parallel. One can understand this as follows. If $\vec t$ effectively points along the $z$-direction, $t_x$ and $t_y$ and thus $\Gamma^\text{so}_\pm$ are zero: the states $\ket{T_\pm}$ are blocked (see Fig.\ \ref{fig:fig3}). As $\ket{T_\pm}$ are eigenstates of the field $B_0$, this blockade could persist up to arbitrarily high fields. Since $\ket{T_0}$ and $\ket{S}$ are rotated into $\ket{\alpha}$ and $\ket{\beta}$, current will then scale in general with the anti-parallel component of spin instead of only the spin singlet.

To conclude, we presented a model to study electron transport in the Pauli spin blockade regime in the presence of strong spin-orbit interaction. It reproduces all features observed in experiment, such as lifting of the spin blockade at high external fields or at low interdot tunnel coupling. We explain the mechanisms involved and identify all relevant energy scales. We also propose a simple way to extend the region of spin blockade.

We acknowledge fruitful discussions with A.\ Pfund, S.\ Nadj-Perge, S.\ Frolov, and K.\ Ensslin. This work is part of the research program of the Stichting FOM.
%, which is financially supported by the `Nederlandse Organisatie voor Wetenschappelijk Onderzoek (NWO)'.


\begin{thebibliography}{99}
\bibitem{endnote1} Throughout the paper we present energies and magnetic fields in terms of frequencies. This corresponds to setting $\hbar = g\mu_B = 1$.
\end{thebibliography}
\end{document}